\documentstyle[12pt,iopconf,psfig]{article}
\def\kms{km~s$^{-1}$ }
\def\Lya{Ly$\alpha$ }

\def\ApJ{{\it Astrophys.~J. }}
\def\AJ{{\it Astron. J. }}
\def\AaA{{\it Astron. Astrophys. }}
\def\MNRAS{{\it Mon. Not. R. Astron. Soc. }}
\def\d{$d_5$ }

\def\cm2{cm$^{-2}$ }
\begin{document}
\title{Measurements of the deuterium abundance in quasar absorption systems}

\author{Scott Burles\footnote{E-mail scott@oddjob.uchicago.edu}} 

\affil{Dept. of Astronomy and Astrophysics,
University of Chicago \\
5640 S. Ellis Ave,
Chicago, IL 60637}

\vspace{-40pt}

\author{David Tytler\footnote{E-mail dtytler@ucsd.edu}} 

\affil{Center for Astronomy and Space Sciences\\
University of California, San Diego\\
9500 Gilman Drive,
La Jolla, CA 92093-0424}

\beginabstract
Observational constraints on the primordial deuterium-to-hydrogen
ratio (D/H) can test theories of the early universe and provide constraints
on models of big bang nucleosynthesis (BBN).  We measure deuterium
absorption in high-redshift, metal-poor QSO absorption systems and directly
infer the value of primordial D/H.  We present two measurements
of D/H, and find D/H = 3.3 $\pm \, 0.3 \times 10^{-5}$
at $z=3.572$ towards QSO 1937-1009 and
D/H = 4.0 $\pm 0.7 \times 10^{-5}$ at $z=2.504$ towards QSO
1009+2956.  Both measurements use multiple-component Voigt profile
analysis of high resolution, high signal-to-noise spectra and
determinations of the Lyman continuum optical depth in low resolution
spectra to constrain the column densities of deuterium and hydrogen.
The measurements are consistent with a single primordial value of
D/H = $3.4 \pm 0.3 \times 10^{-5}$.  This is a
relatively low value, which supports homogeneous models of BBN and
standard models of galactic chemical evolution.
With standard BBN, we find a cosmological baryon-to-photon
ratio, $\eta = 5.1 \pm \, 0.3 \times 10^{-10}$, and a present-day
baryon density in units of the critical density,
$\Omega_b \, h_{100}^2 = 0.019 \pm \, 0.001$.
These values are consistent with high abundance measurements of $^4$He
in extragalactic H~II regions\cite{izo97}, and estimates of $^7$Li in warm, metal-poor
halo stars\cite{bon97}.
\endabstract

\section{The Significance of Deuterium}

In the first one thousand seconds of the universe, 
light nuclei (D, $^3$He, $^4$He, and $^7$Li) are created 
during the epoch of BBN\cite{bbn},\cite{ste98}. 
In the standard model\footnote{with three light neutrino species}
the nuclear yields of the light elements depend on a single 
parameter, $\eta$, the baryon-to-photon ratio.  The density of photons 
in the universe is accurately known from the temperature of the cosmic
microwave background\cite{fix96}.  Therefore, a determination of a single 
primordial light element abundance gives a measure of $\eta$ and
the present-day density of baryons, $\Omega_b$.  The abundance ratio of
deuterium to hydrogen (D/H) is a sensitive function of
$\eta$, and a measurement of primordial D/H places the strong constraints
on $\eta$ and $\Omega_b$.  For example, a measurement of D/H 
with an uncertainty of 20\% corresponds to 12\% uncertainty in both
$\eta$ and $\Omega_b$.

The determination of the primordial abundance ratio of D/H is a 
difficult task.  For the last 25 years, BBN has
stood as the only known cosmological source for deuterium nuclei\cite{ree73}.  
The deuteron is a fragile nucleus, and is easily and totally destroyed through
stellar processing\cite{cla85}.  With no other source of D besides
BBN, any observational measurement of D/H
will always give a lower limit to the primordial ratio.  Local observations,
in either the solar system or interstellar medium, provide a strong lower
bound to the primordial abundance, D/H $> 1.6 \times 10^{-5}$\cite{ism}.
Inferring an upper limit from local measurements of D/H requires
a problematic extrapolation with models of chemical evolution over the last
10 billion years of stellar processing in the galaxy\cite{chemevol}.

In these proceedings, we will describe new measurements of D/H
at high redshift in intergalactic absorption systems 
detected along the line of sight to distant quasars (QSOs).
In 1976, Adams\cite{ada76} realized that high redshift systems showing
strong \Lya absorption would provide an ideal site to determine
the primordial ratio of D/H.  Some of the more compelling reasons
include (1) the high redshift allows much less time for D/H to be
significantly altered from the primordial value, (2) the absorption
systems are very metal poor (less than 1/100 solar
metallicity in most cases), which limits the amount of destruction of D due
to stellar processing, (3) most high redshift absorbers are not likely to
be associated with galaxies, (4) the systems which are useful in a measurement
of D/H must have low temperatures (T $< 20000$ K) and very small
turbulent motions, which limits the amount of energy which could be introduced
into the system.  

\section{The QSO Absorption Systems}

Modern spectroscopy of high redshift QSOs reveal hundreds 
of discrete absorption features which correspond to intervening clouds of
gas which lie along the line of sight.  A small subset of these
absorbers can yield a measurement of D/H, the set defined by the following
criteria:

\begin{itemize}

\item[]{The neutral hydrogen column density (number of atoms per square cm)
must exceed 10$^{17}$ cm$^2$.}

\item[]{The velocity structure along the line of sight must be both
simple and narrow.  Simple, in that only a few components with small
velocity separations are detected in absorption.}

\item[]{The redshift of the absorber must be high, $z_{abs} > 2.5$, to shift
all of the desired absorption features into optical wavelengths which can be
observed by large ground based telescopes.}

\end{itemize}

\noindent{In addition, we prefer systems which show weak or undetectable metal
absorption lines, and lie along the line of sight to the brightest QSOs 
(which makes high-resolution spectroscopic observations practical). }

The criteria listed above places severe limits on the sample of absorption 
systems.  Out of the hundreds of absorbers along the line of sight to
each QSO, we find systems to measure D/H in less then 3\% of the lines of
sight.  That is, less than 0.01\% of the absorbers in the entire
observable sample are suitable systems.  Needless to say, QSO absorption
systems which show deuterium are rare.

The recent success of the high redshift D/H measurements rely 
on a method to increase the likelihood of finding these rare systems.
Systems with the required high H~I column densities are optically thick
to photons with energies above one Rydberg, and exhibit very prominent
continuous absorption in QSO spectra.  These systems are known as
Lyman limit systems. 
Lyman limit systems are easy to detect in low resolution spectra,
which gives a much faster means to find likely systems, compared to obtaining
high resolution data of random lines of sight to look for deuterium absorption.
The top panel of 
Figure 1 shows a low resolution spectrum of QSO 1937--1009.  The Lyman
limit system at $z=3.572$ is responsible for the  
continuous absorption of flux below 4200 \AA.  The Lyman limit serves
not only to identify the systems, but the optical depth of the absorption
is directly proportional to the H~I column density in the system.  The
role of the Lyman limit measurements will be discussed in detail in the
next section.

\begin{figure}
\psfig{file=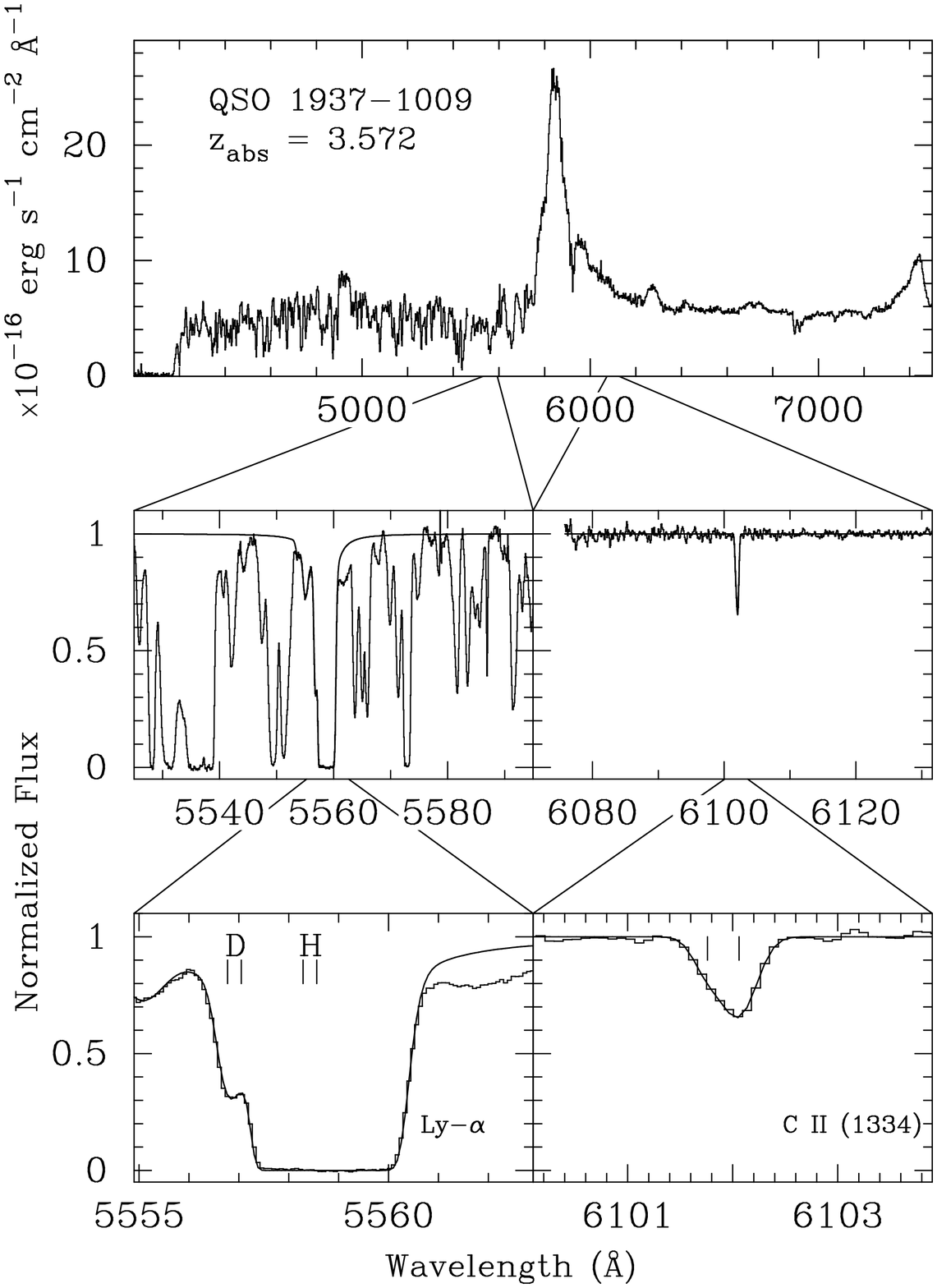,width=\textwidth}
\caption{QSO 1937-1009: See text for more details
\protect\\
{\it Top:} Lick spectrum (FWHM = 4 \AA)
\protect\\
{\it Middle and Bottom:} Keck+HIRES spectrum of the \Lya region (left)
and C~II region (right) at $z=3.5722$ }
\end{figure}

Figure 1 shows a schematic view of two absorption features 
in the deuterium system towards QSO 1937--1009.  The low resolution
spectrum in the top panel covers a large spectral range and is
plotted in units of flux per \AA.  The spectrum shows the intrinsic shape
of the QSO continuum, including emission features, but also depicts the
sharp increase in absorption at wavelengths shorter than \Lya emission
at 5850 \AA, due to the intervening \Lya absorption systems 
(the ``\Lya forest'').
The next set of panels zooms in on two specific regions in the spectrum,
corresponding to the wavelengths of \Lya and singly ionized carbon (C~II)
redshifted to $z=3.572$ ($\lambda = \lambda_0 (1 + z)$).
The two regions are noticeably different.  The left hand panel shows the
high density of lines in the \Lya forest at high redshift surrounding the
strong \Lya line of the deuterium system.  The right hand panel shows
only the weak optically thin absorption from C~II,
and shows the absence of any other features nearby, so there is no confusion.
The bottom panels show the profiles of the \Lya and C~II.
Deuterium absorption can be see on the left wing of the H~I \Lya feature.
The isotopic separation of D and H lines is -82 km s$^{-1}$; the D~I
energy levels are shifted up in energy by a factor of 1.000272 compared
to hydrogen, due to the small change in the reduced mass from the additional
neutron.
The middle and bottom panels show spectra obtained with the high resolution
echelle spectrograph (HIRES)\cite{vog94}.  The spectra have a resolution
of 8 \kms FWHM, which is sufficient to resolve the deuterium lines.
The spectra have been normalized to allow a direct analysis of the
absorption features.  The intrinsic QSO continuum was modeled with a
low order polynomial in each region. 
The next section will describe the method to measure D/H from the Lyman
series absorption profiles of D~I and H~I.

\section{The Deuterium Measurements}

Here we present the spectra and analysis of the two absorption systems
which yield the best measurements of D/H at high redshift.
For a more detailed description of the method and analysis,
the reader is referred to larger papers\cite{bur98a,bur98b,bur97,tyt96,tyt97}.
A brief description is given here.

\subsection{The Method to Measure D/H}

We construct a model fit with discrete absorption lines,
given by Voigt profiles,
convolved with the instrumental resolution.  Each Voigt profile is
given by three parameters, the column density (N), the redshift ($z$),
and the intrinsic velocity dispersion along the line of sight ($b$).
Each hydrogen line profile with a large hydrogen column, 
N(H~I) $> 10^{16}$ \cm2 has a corresponding deuterium profile at the
same redshift.  The velocity dispersion of both the H~I and D~I profile
are given by two parameters, the temperature and turbulent velocity dispersion.
In short, we substitute the free parameters of b(H~I) and b(D~I) with
two parameters, T and $b_{tur}$, which are identical in both H~I and
D~I absorption lines.  We also assume that D/H is global, that is,
all components have the same ratio of N(D~I)/N(H~I).

\begin{figure}
\psfig{file=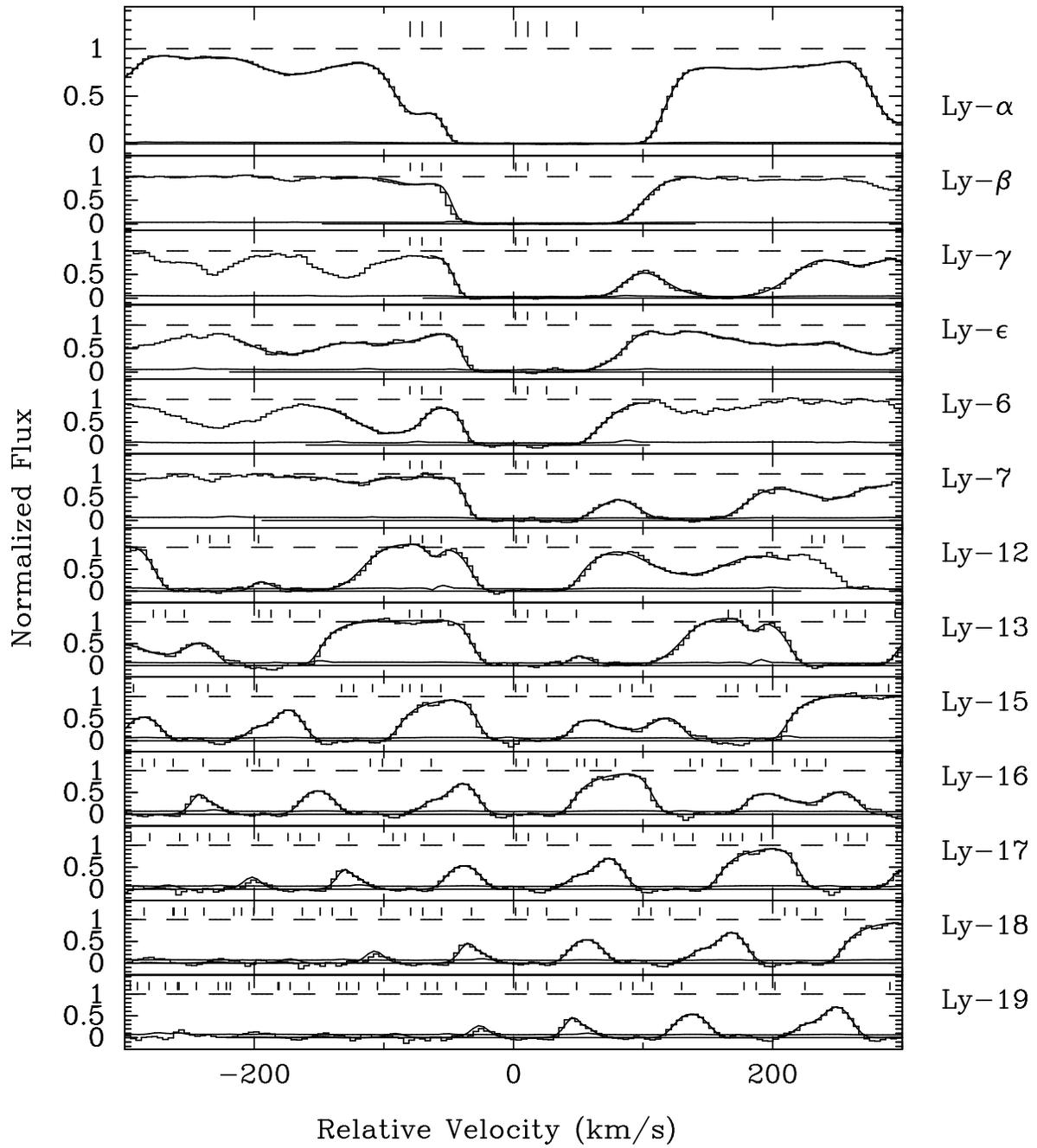,width=\textwidth}
\caption{Lyman series of Q1937-1009 at $z=3.5722$.}
\end{figure}

\begin{figure}
\psfig{file=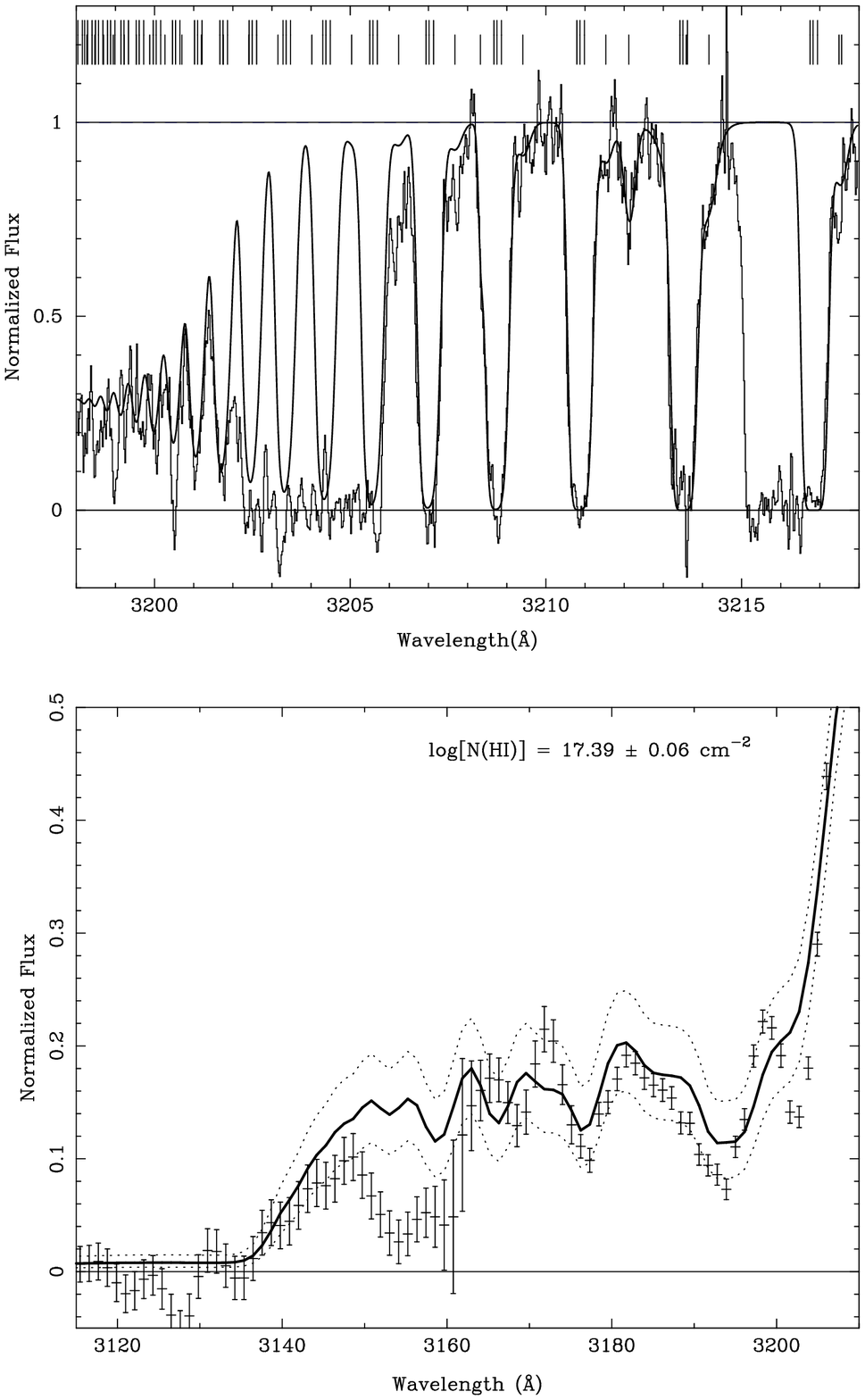,width=\textwidth}
\caption{Lyman limit region of Q1009+2956 at $z=2.504$.
\protect\\
{\it Top}: HIRES spectrum (FWHM = 8 km/s) showing lines Ly-11 to Ly-24.
{\it Bottom}: Lick spectrum (FWHM = 4 \AA) and model
fit short-ward of Lyman limit.  The solid line shows the best fit 
and dotted lines show 1$\sigma$ errors.}
\end{figure}

We select regions of the HIRES spectrum which contain the absorption
features of the Lyman series and are not severely blended by 
unrelated lines in the \Lya forest.  The continuum is modeled as a low
order polynomial in each region, and the coefficients of the polynomial
are treated as free parameters in the fitting process to statistically
allow for continuum uncertainties.  

In this analysis, a measure of D/H is the goal, so we study the
goodness of fit $\chi^2$ as a function of D/H.  To directly study the
effect of D/H on the model fit, we model the spectra with hundreds
of iterations.  In each iteration, we specify D/H, let all other parameters
go free, and find the best $\chi^2$ associated with the specified 
value of D/H.  We iterate over a large range in D/H, and measure the most
likely value of D/H by locating the minimum of $\chi^2$ as a function
of D/H.  This method has much to offer.  The parameters describing the
profiles and continuum are sometimes highly correlated.  The true uncertainty
in D/H can not be determined by calculating the individual parameter
uncertainties near the minimum of the $\chi^2$ function.  We must explicitly
map the $\chi^2$ dependence on D/H to include the correlations which are
present in the model.

\subsection{QSO 1937-1009}

Tytler \etal (1996)\cite{tyt96}
made the first measurement of low D/H in the absorption
system at $z=3.572$ towards Q1937--1009.  We analyzed the 
high-resolution spectrum (8 hrs of exposure), which resolved the entire Lyman
series up to Ly-19, as well as associated metal lines.  By profile
fitting the Lyman lines, with the position of the velocity components given
by the metal lines, we find D/H = 2.3 $\pm \, 0.3 \pm 0.3 \times 10^{-5}$
(statistical and systematic errors).  
The largest uncertainty in the measurement is the neutral hydrogen column
density, log N(H~I) = 17.94 $\pm \, 0.06 \pm 0.05$, 
and the uncertainty stems from the saturated Lyman profiles
(discussed in detail below).  We then obtained a high quality 
low-resolution spectra from Keck with LRIS\cite{oke95}, which gave better
sensitivity short-ward of the Lyman limit, to directly measure the
total N(H~I) in the system and therefore place better constraints on D/H.
Utilizing both the high and low-resolution spectra, we find
log N(H~I) = 17.86 $\pm \, 0.02$ by a direct measurement of the optical depth
short-ward of the Lyman limit at 4200 \AA~(Burles \& Tytler 1997a).
With this constraint and the more sophisticated method described above, 
we measure D/H = 3.3 $\pm \, 0.3 \times 10^{-5}$ (Burles \& Tytler 1997b).

\begin{figure}
\psfig{file=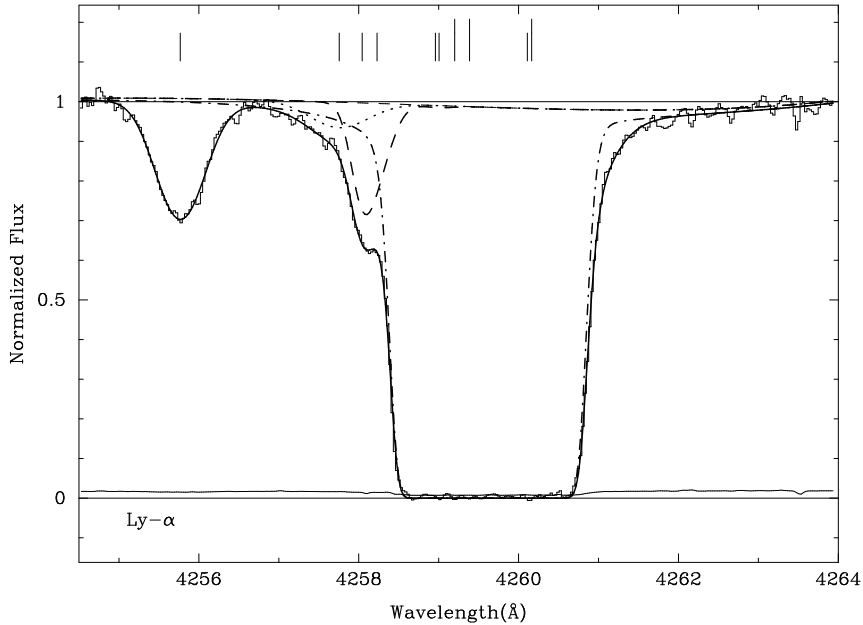,width=5.2in}
\caption{ 
\Lya region of $z=2.504$ system towards Q1009+2956. 
The histogram shows the pixels of the HIRES spectrum normalized to the
initial continuum estimate (solid line at unity).  
The model profile is composed of the main H~I (dot-dashed) and D~I (dashed), 
and contaminating H~I (dotted). }
\end{figure} 

The Lyman lines used in the fitting procedure are shown in Figure 2.
The spectra have been shifted to a velocity scale by the simple
transformation, 
${{v} \over {c}} = {{\lambda} \over {\lambda_0  (1 + z)}} - 1.0,$
where $\lambda_0$ is the rest wavelength of the Lyman line, and $z$
is the redshift corresponding to zero velocity, in this case $z=3.57221$.

Although the positions of the metal line components were previously
used to constrain the fit\cite{tyt96}, the present analysis does not
require any information from the metal lines in the fitting procedure.
The use of the metal lines can be justified for this system, but in
general, one is likely to introduce systematics by including the metal lines,
and it remains a good policy to leave all other lines except H and D out
of the fitting procedure.

\section{QSO 1009+2946}
We discovered an absorption system at $z=2.504$ towards 
Q1009+2956 ideal for a measurement of D/H\cite{tyt97}.  
This system has a lower
hydrogen column density, log N(H~I) = $17.39 \pm 0.06$.  The
highest order Lyman lines become unsaturated, which yields a precise 
measurement of N(H~I) in both low and high resolution spectra (Fig. 3). 
Over twelve hours of Keck+HIRES produced a very high quality spectrum
of the entire Lyman series, resolving the entire series up to Ly-22.
We find strong evidence for contamination of the deuterium \Lya absorption
feature, which introduces the largest uncertainty in the
measurement.  Figure 4 shows the \Lya line responsible for the
contamination.  The profiles of D~I (dashed lines) and H~I (dash-dotted lines)
are shown explicitly, with the weak contaminating hydrogen line (dotted)
falling just blueward (to the left) of D~I.  We include this contamination
as a free parameter and find 
D/H = 4.0 $\pm \, 0.7 \times 10^{-5}$\cite{bur98b}.

\section{Conclusions}

These two measurements of D/H in QSO absorption systems are the best
and most robust measures to date.  Deuterium has been identified
and analyzed in a number of other QSO absorption systems\cite{dhother}
We have found another two systems which place a strong upper limit on D/H
at D/H $< 10^{-4}$.  Combined with the two measurements described above,
the four independent systems support a low primordial abundance of deuterium,
and together give D/H = 3.4 $\pm \, 0.3 \times 10^{-5}$.  
If this represents the primordial value, nucleosynthesis calculations
from standard BBN models with three light neutrinos give
$\eta =  5.1 \pm \, 0.3 \times 10^{-10}$ and
$\Omega_b\,h_{100}^2 = 0.019 \pm \, 0.001$.

The constraints from D/H can be utilized to constrain cosmological models,
quantify dark matter both in unobserved baryons and non-baryons, specify
the zero point for models of deuterium evolution\cite{chemevol}, 
test directly the
predictions of standard BBN by comparing with other light element
abundances\cite{hat}, 
and limit the amount of small scale entropy fluctuations
in the early universe\cite{jed}.

\section*{Acknowledgements}
SB is indebted to David N. Schramm for the support and encouragement
to both present and participate at this symposium.  
We thank George Fuller, Karsten Jedamzik, David Kirkman, Martin Lemoine,
Jason X. Prochaska and Michael S. Turner for many useful converations.
We would like to thank Tony Mezzacappa and
the symposium staff for their effort and hospitality.
This work was supported by NASA grant NAG5-3237 and NSF grant AST-9420443.


\vspace{-14pt}
\normalsize
\begin{thebibliography}{99}

\bibitem{izo97}
Izotov Y I, Thuan T X and Lipovetsky V A 1997 \ApJ {\it Sup.} {\bf 108} 1

\bibitem{bon97}
Bonifacio P and Molaro P 1997 \MNRAS {\bf 285} 847

\bibitem{bbn}
Walker T P, Steigman G, Schramm D N, Olive K A and Kang H S 1991
\ApJ {\bf 376} 51
\newref
Smith M S, Kawano L H and Malaney R A 1993 \ApJ {\it Sup.} {\bf 85} 219
\newref
Copi C J,Schramm D N and Turner M S 1995 {\it Science} {\bf 267} 192
\newref
Sarkar S 1996 {\it Rep Prog Phys} {\bf 59} 1493

\bibitem{ste98}
Steigman G 1998, these proceedings

\bibitem{fix96}
Fixsen D J, Cheng E S, Gales J M, Mather J C,
Shafer R A and Wright E L 1996 \ApJ {\bf 473} 576

\bibitem{ree73}
Reeves H, Audouze J, Fowler W A and Schramm D N 1973
\ApJ {\bf 179} 909

\bibitem{cla85}
Clayton D D 1985 \ApJ {\bf 290} 428

\bibitem{ism}
Griffin \etal 1996 \AaA {\bf 315} L389
\newref 
Piskunov N, Wood B E, Linsky J L, Dempsey R C and Ayres T R 1997
\ApJ {\bf 474} 315
\newref Gautier D and Morel P 1997 \AaA {\bf 323} L9
\newref Chengalur J N, Braun R and Butler B W 1997 \AaA {\bf 318} L35

\bibitem{chemevol}
Steigman G and Monica T 1995 \ApJ {\bf 453} 173
\newref Fields B 1996 \ApJ {\bf 456} 478
\newref
Copi C J 1997 \ApJ {\bf 487} 704
\newref Timmes F X, Truran J W, Lauroesch J T and 
York D G 1997 \ApJ {\bf 476} 464

\bibitem{ada76}
Adams T F 1976 \AaA {\bf 50} 461
 
\bibitem{vog94}
Vogt S \etal 1994 {\it Proc SPIE} {\bf 2198} 362

\bibitem{bur98a}
Burles S and Tytler D 1997 \ApJ in press (astro-ph 9712108)

\bibitem{bur98b}
Burles S and Tytler D 1997 submitted to \ApJ (astro-ph 9712109)

\bibitem{bur97}
Burles S and Tytler D 1997 \AJ {\bf 114} 1330

\bibitem{tyt96}
Tytler D, Fan X-M and Burles S 1996 {\it Nature} {\bf 381} 207 

\bibitem{tyt97}
Tytler D and Burles S 1997 {\it Origin of Matter and
Evolution of Galaxies} eds.  Kajino T, Yoshii Y and Kubono S
(World Scientific Publ Co: Singapore) p 37

\bibitem{oke95}
Oke J B, Cohen J G, Carr M, Cromer J, 
Dingizian A, Harris F H, Labrecque S, Lucinio R, 
Schaal W, Epps H and Miller J 1995 {\it Pub. Astron. Soc. Pac} {\bf 107} 375

\bibitem{dhother}
Songaila A, Cowie L L, Hogan C J and Rugers M 1994 {\it Nature} 
{\bf 368} 599
\newref Carswell R F, Rauch M, Weymann R J, Cooke A J and Webb J K
1994 \MNRAS {\bf 268} L1
\newref Carswell R F, Webb J K, Lanzetta K M, Baldwin J A, Cooke A J,
Williger G M, Rauch M, Irwin M J, Robertson J G and Shaver P A
1996 \MNRAS {\bf 278} 506
\newref Wampler E J, Williger G M, Baldwin J A, Carswell R F,
Hazard C and McMahon R G 1996 \AaA {\bf 316} 33
\newref Rugers M and Hogan C 1996 \ApJ {\bf 459} L1
\newref Rugers M and Hogan C 1996 \AJ {\bf 111} 2135
\newref 
Webb J K, Carswell R F, Lanzetta K M, Ferlet R, Lemoine M,
Vidal-Madjar A and Bowen D V 1997 {\it Nature} {\bf 388} 250

\bibitem{hat}
Cardall C Y and Fuller G M 1996 \ApJ {\bf 472} 435
\newref Fuller G M and Cardall C Y 1996 {\it Nucl Phys B} {\bf 51} 71
\newref 
Hata N, Steigman G, Bludman S and Langacker P 1997 {\it Phys Rev D}
{\bf 55} 540
\newref Schramm D N and Turner M S 1997 submitted to {\it Rev. Mod. Phys.}
(astro-ph 9706069)

\bibitem{jed}
Jedamzik K 1993 {\it Ph.D. Thesis} University of California, San Diego
\newref Jedamzik K and Fuller G M 1994 \ApJ {\bf 423} 33;
1995 \ApJ {\bf 452} 33
\newref Jedamzik K, Fuller G M and Mathews G 1994 \ApJ {\bf 423} 50

\end{thebibliography}
\end{document}